\documentstyle[11pt]{article}
\begin{document}

\begin{flushright}
BRX TH-510
\end{flushright}

\begin{center}

{\large\bf Some Remarks on Dirac's Contributions to General
Relativity}

S. DESER

Brandeis University \\
Department of Physics \\
Waltham, MA 02454, USA\\
E-mail: deser@brandeis.edu

\end{center}



{\abstract I provide a very brief sketch of some of Dirac's
interests and work in gravity, particularly his Hamiltonian
formulation of Einstein's theory and its relation to his earlier
research.}

It is a great honor for any theoretical physicist to speak at a
Memorial for Professor Dirac (like many others, I cannot bring
myself to call him Paul!), whom I last saw when I gave a
Colloquium here in Tallahassee shortly before he died. He went
through the canonical behavior of sleeping during my talk, then
awakening at the end with a perfectly reasonable question. He had
expressed a desire to see me afterwards in his office; there, he
immediately inquired what was new in physics. I told him that he
probably wouldn't be pleased to hear that there had appeared a
finite quantum field theory, whereupon he asked -- unhappily --
what it was. When told it was $N$=4 supersymmetric Yang--Mills, he
denied knowledge of any of those words.  After hearing a suitable
translation, his reaction was simply that either there was an
error in the calculations or that the model was really
non-interacting, an opinion then also held by some pros. It was
clear that his mind was made up about the bankruptcy of QFT, but
then again, he was also one of the inventors of extended objects.

It is {\it de rigueur} to include a Dirac story in any lecture of
which he is the subject, so that by now there are very few unknown
ones. My favorite, because physically fraught, comes from Abdus
Salam's introduction to Dirac's talk at the famous 1968 Trieste
Conference. When Dirac first came to St. John's, there was a
traditional Christmas event at which the Maths tutors would pose a
riddle to the incoming students; here's what he drew: Three
exhausted fishermen wash up on the beach with their haul, but are
too sleepy to split it.  At dawn, the first fisherman awakens,
takes his 1/3 share and goes home,  throwing the one leftover fish
back into the sea. Next, number two, unaware of the first, acts
identically, as does, finally the third one. What is the minimum
(integer) size $N$ of the original pile? Dirac's lightning-fast
solution: $N$=--2\,!  Not only does it foreshadow that other Dirac
sea, but displays a fixed point invariance besides -- every
fisherman sees the {\it same} pile! [To keep my audience from
calculating, here is the equation: $4N = 9P + 10$ where $P$ is the
pile faced by the late sleeper. Of course, $N = 25$ is the
``correct", and so much duller, answer.]

What is {\it not} needed from me is an encomium: Dirac was a true
Martian, a Hungarian\linebreak (-in-law) one at that. Feynman and
Schwinger, neither otherwise overly impressionable, regarded him
with awe, and he was right near Einstein and Bohr on Landau's
famous logarithmic rankings. Instead, in the short time available,
I will give a brief (but necessarily incomplete) appreciation of
some of Dirac's work in General Relativity (GR), which I was in a
position to observe. Here, I only cite the original papers, but
not later lectures or reprises.\footnote{There exists \cite{001} a
very useful ``Dalitz plot", referencing Dirac's complete works,
and reprinting those published before 1949.}

Let me begin by setting the historical stage.  After its rapid
initial successes, GR was very much a stepchild of theoretical
physics research for three decades, until the early
fifties.\footnote{The first, and completely isolated, attempts at
treating (linearized) GR as a dynamical system were probably those
of the early thirties by Rosenfeld \cite{002}, and by Bronstein
\cite{003} in the USSR. Tragically, the latter's name coincided
with Trotsky's real one, and he disappeared early in the Stalin
purges.} Even then, the renaissance to which Dirac's work belonged
was primarily disconnected from the (indifferent or hostile) field
and particle theory mainstream. Indeed, this separation was
traditional.  While GR was understood quickly after its discovery,
neither Bohr nor Heisenberg, for example, ever ventured there,
although the former did use the equivalence principle against
Einstein in a famous debate on quantum mechanics at the 1927
Solvay Conference, and the latter was the first, in the late
thirties, to understand why perturbative quantization of theories,
with positive dimensional (self-)coupling constant would fail.
Pauli of course started life writing a text on GR, but despite
continued interest, he never really contributed to it at the
``Pauli" level. In later years, Schr\"{o}dinger did venture into
the field with some brilliant pedagogical expositions, but alas
mostly into the morass of ``unification by nonsymmetic metric"
that occupied Einstein's own late years. Born explicitly wrote
that once he understood GR, he vowed never to work on it. Thus
(apart perhaps from Jordan and Klein), Dirac was unique among the
creators of quantum mechanics to work seriously on GR.

Looking at Dirac's own earlier work, one is struck first by some
seemingly disconnected, but indicative, themes.  The first,
physics in deSitter (dS) space \cite{004}, comes (as usual) out of
nowhere. There he argued that ``masslessness" means a
non-vanishing mass parameter for spin 1/2 (see also \cite{005}),
to some extent foreshadowing the spin 3/2 properties of
cosmological supergravity \cite{006,007}. [Amusingly, Dirac worked
only in dS rather than in AdS, and so required this spin 1/2 mass
to be imaginary, $m\sim \sqrt{-\Lambda}$, instead of real as in
the natural SUGRA domain, AdS\,!] More formally, Dirac exploited
the Weitzenbock identity for gravity, just as he first did to
discover the $g=2$ factor for the Dirac electron in a background
magnetic field. To analyze propagation, one must first square the
Dirac equation to get a wave operator.  When the partial
derivative is replaced by a covariant one, the square is more
complicated: apart from factors of 1/2, $i$ etc., one has
schematically
\begin{equation}
D \!\!\!\! / \, ^2 \equiv \frac{1}{2} \{ \gamma ,\gamma \} DD +
\frac{1}{4} [\gamma , \gamma ][D,D]\; .
\end{equation}
The last term vanishes for ordinary derivatives, but in general
the commutator $[D,D]$ defines a curvature (``field strength") and
in the gravitational case, its net effect is to become the scalar
curvature, which (in AdS) is proportional to the cosmological
constant $\Lambda$. Consequently,
\begin{equation}
(D \!\!\!\! / + m)(D \!\!\!\! / - m) = D^2 - (\Lambda + m^2)
\end{equation}
and propagation seems to be on the light cone only if $\Lambda +
m^2 =0$; actually for s=1/2, things are a bit more subtle and
$m$=0 is in fact the counterpart of the conformally improved
scalar there. For higher (also integer) spins, things get even
more interesting, as described in \cite{008}. Dirac was to return
to this theme, but it really belongs to QFT in a (constant
curvature) gravitational background rather than to dynamic
gravity.

A more direct lead to GR is Dirac's abiding interest in
Lagrangian/Hamiltonian dynamics, locality and ``unusual" systems.
Two \cite{009,010} of the great articles on formulations of
dynamical systems are well-known.  However, there is one other
which is significant because he did {\it not} use it, namely
introduction \cite{011} of what Dirac calls ``homogenous
coordinates". This is nothing but a variant of the Jacobi form of
the traditional action principle for normal systems in flat space
in which time and Hamiltonian form a new, ``$n$ + first",
conjugate $(q,p)$ pair. In terms of the extended set, and of the
Lagrange multiplier $\lambda$ that keeps the ``true number" of
degrees of freedom to be the original $n$,  the initial Lagrangian
$$
L = \sum^n_{i=1} p_i\dot{q}_i - H(p,q) \eqno{(3\rm{a})}
$$
becomes
$$
 L = \sum^{n+1}_{i=1} p_i\dot{q}_i + \lambda \tilde{H} (p,q) \;.
 \eqno{(3\rm{b})}
$$
where the constraint $\tilde{H} = 0$ has a root $p_{n+1} = -
H(p,q)$. As it happens, in our (ADM) contemporaneous and
independent development of Einstein theory \cite{012},
understanding that the Einstein--Hilbert (or any other
diffeo-invariant) action was necessarily an ``already
parametrized" system {\it a la} (3b) was an essential, beautiful,
confirmation of our canonical formulation: The big difference from
the traditional Jacobi applications is that there is no {\it a
priori} passage in GR back from (3b) to (3a): indeed, there is no
(3a)! Instead one could fix a gauge, $q_{n+1} = t$ and solve the
constraint for its conjugate $p_{n+1}$ almost arbitrarily (within
physical bounds). [We learned about Jacobi from the Lanczos book
on variational principles, rather than from \cite{011}.] In any
case, looking back, it is hard to understand why Dirac never
explicitly invoked the Jacobi formulation for GR, given that one
of his motivations was surely to apply his general formalisms to
this challenging system!

Dirac's results were concentrated in three papers
\cite{013,014,015}.\footnote{The density of competing groups in
the field was so low that the dilute approximation can be used in
assessing Dirac's contribution, which like ours was also foreign
to that of most relativists. For an idea of the changing research
directions that characterized the era, the Proceedings \cite{016}
of several conferences bracketing this time are instructive: the
first, in Bern, was in 1955, then came Chapel Hill in 1957,
Royaumont in 1959 and Warsaw in 1962.} Let us briefly summarize
the salient points, starting from  Dirac's realization that a 3+1
decomposition of the gravitational field variables (rather than
maintaining manifest 4-covariance) is essential to any Hamiltonian
-- and hence quantization -- description. He accordingly used
projections with respect to a $t$=const.\ surface and noted that
the metric components $g_{\mu 0}$ are only lapses and shifts (as
they were later called) and not dynamical. He thus inferred that
the time development of any dynamical ({\it i.e.}, not involving
$g_{\mu 0}$) gravitational variable $\eta$ is governed by an
evolution equation of the form
$$
\dot{\eta} = \int d^3x [N\xi_L + N_i \xi^i ] \; , \;\;\; N\equiv
(-g^{00})^{-1/2} \; , \;\;\; N_i \equiv g_{0i}  \eqno{(4)}
$$
and hence must be derived from a Hamiltonian
$$
H = \int d^3x (N H_L + N_i H^i) \; ; \;\;\; H_L \equiv \, ^3\!\!R
+ (p^2_{ij} - p^2_{ij}/2) \; , \;\;\; H^i = -D_j p^{ij} \eqno{(5)}
$$
where $H_L,H^i$ depend as shown on the conjugate (but {\it not}
independent) pairs of spatial tensors $(g_{ij},p^{ij})$, which
(``weakly") obey the standard set of Poisson brackets.  Here
$^3\!\!R$ is the intrinsic 3-space curvature scalar and $p^{ij}$
is essentially the second fundamental form. The $(H_L, H^i)$ are
four (``weak" or ``secondary") constraints on the $(g_i, p^{ij})$,
and they are understood to be combinations of the $G^0_\mu$
components of the Einstein tensor, because those do not depend on
second time derivatives.  That property is easily noted from the
Bianchi identities, $\partial_0 G_\mu^0 \sim (\partial_i + \Gamma
) G$: the right side has only second time derivatives since the
$G_{\mu\nu}$ there are only spatially differentiated, so the left,
$G^0_\mu$, must only have first time derivatives since {\it it} is
time-differentiated. At this point, the natural -- but hardly
unique -- coordinate choice $g_{\mu 0} = - \delta_{\mu 0}$ is
invoked to yield the ``true" Hamiltonian $H_{\rm MAIN}$, which is
just $\int d^3x {\cal H}_L$ after dropping the total 3-divergence
that is the linear part of the 3-scalar curvature, {\it i.e.}, the
(ADM) energy. [Dirac's third work \cite{015} is concerned with
establishing the form of the field's energy, and evaluate it for
some special cases.] Dirac then went on to discuss general issues
of constrained Hamiltonian dynamics, in the spirit of
\cite{009,010}, in order to understand coordinate fixing and
reduction of the apparent number of variable pairs by the
constraints.  The choice of time is the obvious minimal surface
one, $p^i_i = 0$, leaving 5 pairs and finally a spatial harmonic
gauge is invoked for the remainder, and the resulting P.B.\ are
then discussed in a perturbative way. The second paper ends with a
short section entitled ``Quantization", which simply says that
since a complete set of commuting variables is formed by the
unimodular, divergenceless part of the spatial metric (together
with matter variables), the wave function(al)  that depends only
on this reduced space is unconstrained, and thus obeys a simple
Schr\"{o}dinger equation.  As can be expected, Dirac worries about
quantum problems: he notes that the $t$=const.\ surfaces must
remain spacelike, which means that $g_{ij}$ has to maintain a
positive signature -- {\it i.e.}, positive determinant. Since this
quantity is for him also related to the ``energy density", he
states that violation could occur very near point sources,
basically due to negative gravitational self-energy.  The
concluding sentence is: ``The gravitational treatment of point
particles thus brings in one further difficulty, in addition to
the usual ones in the quantum theory." This is rather curious coda
since the above problems are really as relevant classically, and
of course they are very different from the perturbative
nonrenormalizability issues that have dominated all subsequent
studies. After this pioneering foray, Dirac's original
publications in the field waned, apart from one later paper
\cite{017} on conformally invariant extensions of GR.

I close by emphasizing that, in an era when geometry was (for
``real" relativists) the vital guiding thread to the mysteries of
GR, the Hamiltonian approaches provided an alternate framework,
common to nonabelian gauge QFT, in which the gravitational field
is regarded as a regular dynamical system with degrees of freedom,
asymptotic boundary conditions and global conserved quantities
correlated to the chosen asymptotics.  The Hamiltonian approach
also serves well when the cosmological constant does not vanish,
not to mention its role in elucidating supergravity, which is
rightly described as the ``Dirac square root" of Einstein theory!
Although Dirac lived to see both supergravity and (pre-revolution)
string theory, he was, like Moses, unable to enter the promised
land, to which he had been our guide in so many ways. We all stand
on Dirac's shoulders.

\end{document}